\documentclass[12pt,a4paper]{article}
\pdfoutput=1

\usepackage{geometry}
\usepackage{amsmath}
\usepackage{amssymb}
\usepackage[dvips]{graphicx}
\usepackage{cite}
\usepackage{pstricks}
\usepackage{bm}
\usepackage{pbox}
\usepackage{placeins}
\usepackage{graphicx}
\usepackage{caption}
\usepackage{subcaption}
\usepackage[T1]{fontenc}
\usepackage{footnote}
\usepackage{pdfpages}
\usepackage{hhline}
\usepackage{multirow}
\usepackage{enumitem}
\usepackage[toc,page]{appendix}
\usepackage{dsfont}
\allowdisplaybreaks

\definecolor{MyDarkBlue}{rgb}{0.1, 0.1, 0.8} 
\definecolor{MyLightBlue}{rgb}{0.22,0.51,0.9}
\definecolor{MyGreen}{rgb}{0.0, 0.5, 0.0}
\definecolor{BrickRed}{rgb}{0.8, 0.25, 0.33}
\usepackage[colorlinks=true,linkcolor=blue,citecolor=MyDarkBlue,
urlcolor=MyLightBlue,bookmarksnumbered=true,bookmarksopen]{hyperref}
\hypersetup{colorlinks, citecolor=BrickRed,linkcolor=MyDarkBlue, urlcolor=MyLightBlue}

\begin{document}
\vspace*{-0.2in}
\begin{flushright}
OSU-HEP-19-08
\end{flushright}
\vspace{0.5cm}
\begin{center}
{\Large\bf 
Towards Minimal $SU(5)$
}\\
\end{center}

\vspace{0.5cm}
\renewcommand{\thefootnote}{\fnsymbol{footnote}}
\begin{center}
{\large
{}~\textbf{Ilja Dor\v{s}ner}$^{a,b,}$\footnote{ E-mail: \textcolor{MyLightBlue}{dorsner@fesb.hr}} and 
{}~\textbf{Shaikh Saad}$^{c,}$\footnote{ E-mail: \textcolor{MyLightBlue}{shaikh.saad@okstate.edu}}
}
\vspace{0.5cm}

$^{a}${\em  University of Split, Faculty of Electrical Engineering, Mechanical Engineering and Naval Architecture in Split (FESB), Ru\dj era Bo\v{s}kovi\'{c}a 32, 21000 Split, Croatia}\\
$^{b}${\em  Jo\v{z}ef Stefan Institute, Jamova 39, 1000 Ljubljana, Slovenia}\\
$^{c}${\em Department of Physics, Oklahoma State University, Stillwater, OK 74078, USA }
\end{center}

\renewcommand{\thefootnote}{\arabic{footnote}}
\setcounter{footnote}{0}
\thispagestyle{empty}
\begin{abstract}
We propose a simple $SU(5)$ model that connects the neutrino mass generation mechanism to the observed disparity between the masses of charged leptons and down-type quarks. The model is built out of $5$-, $10$-, $15$-, $24$-, and $35$-dimensional representations of $SU(5)$ and comprises two (three) $3 \times 3$ ($3 \times 1$) Yukawa coupling matrices to accommodate all experimentally measured fermion masses and mixing parameters. The gauge coupling unification considerations, coupled with phenomenological constraints inferred from experiments that probe neutrino masses and mixing parameters and/or look for proton decay, fix all relevant mass scales of the model. The proposed scenario places several multiplets at the scales potentially accessible at the LHC and future colliders and correlates this feature with the gauge-boson-mediated proton decay signatures. It also predicts that one neutrino is massless. 
\end{abstract}

\newpage
\setcounter{footnote}{0}
\section{Introduction}
\label{SEC-01}

The promising proposal of unification of the Standard Model (SM) matter fields and their interactions using $SU(5)$ group as the supporting gauge structure has been around for more than four decades~\cite{Georgi:1974sy}. The initial efforts, naturally, have been devoted to understanding of what turned out to be a finite number of possible ways to generate masses for the SM charged fermions~\cite{Georgi:1979df,Ellis:1979fg}. The subsequent need to accommodate the neutrino mass generation mechanism, on the other hand, has revealed that there are many potentially viable paths that could be taken and the majority of the studies within the $SU(5)$ theory framework, over the last two decades, has been focused on various ways of implementing it~\cite{Dorsner:2005fq, Bajc:2006ia, Dorsner:2006fx, Perez:2007rm, Bajc:2007zf, Dorsner:2007fy, Schnitter:2012bz, DiLuzio:2013dda, Dorsner:2014wva, Tsuyuki:2014xja, Perez:2016qbo, Hagedorn:2016dze, Babu:2016aro, Dorsner:2017wwn, Kumericki:2017sfc, FileviezPerez:2018dyf, Saad:2019vjo, Klein:2019jgb}.

In this manuscript, we combine several known mechanisms of mass generation for both the charged and neutral fermions of the SM within the $SU(5)$ framework to produce an unexpectedly economical and predictive model that might take us a step closer towards a minimal $SU(5)$. The scalar sector of our proposal contains only three representations, whereas the fermion sector, besides the SM fields, contains only one vectorlike representation. In addition to that the proposal allows for only two (three) $3 \times 3$ ($1 \times 3$) Yukawa coupling matrices that need to accommodate all experimentally measured fermion masses and mixing parameters, where one of the $3 \times 3$ matrices can be taken, without the loss of generality, to be diagonal and real whereas the other $3 \times 3$ matrix is symmetric. Finally, the gauge coupling unification analysis, coupled with the need to accommodate experimentally observed parameters in the neutrino sector, constrains all relevant mass scales of the model to reside within very narrow ranges of viability.

The manuscript is organized as follows. We specify the particle content of our proposal in Sec.~\ref{SEC-02}. There we also discuss the particularities of the gauge coupling unification, spell out the connection between different mass generation mechanisms, and present numerical study that showcases the phenomenological viability of the proposal. We conclude in Sec.~\ref{SEC-03}.

\section{Proposal}
\label{SEC-02}

\subsection{Particle content}

We propose an extension of the original Georgi-Glashow (GG) model that connects the neutrino mass generation to the observed disparity between the masses of charged leptons and down-type quarks. The scalar sector of the GG model comprises one $5$-dimensional and one $24$-dimensional representation. We add to it a single $35$-dimensional representation. The decompositions of these representations under the SM gauge group $SU(3) \times SU(2) \times U(1)$ are $5_H\equiv \Lambda=\Lambda_1(1,2,1/2)+\Lambda_3(3,1,-1/3)$, $24_H\equiv \phi=\phi_0(1,1,0)+\phi_1(1,3,0)+\phi_3(3,2,-5/6)+\phi_{\overline{3}}(\overline{3},2,5/6)+\phi_8(8,1,0)$, and $35_H\equiv \Phi=\Phi_1(1,4,-3/2)+\Phi_3(\overline{3},3,-2/3)+\Phi_6(\overline{6},2,1/6)+\Phi_{10}(\overline{10},1,1)$, where the vacuum expectation value (VEV) of $24_H$ breaks $SU(5)$ down to the SM gauge group. Our normalisation for the VEV of $24_H$ is such that $\langle 24_H \rangle \equiv \langle
\phi \rangle=v_{24}/\sqrt{15} \ \textrm{diag}(-1,-1,-1,3/2,3/2)$. The electroweak VEV resides in $5_H$ and we denote it with $v_H$, where $v_H=174$\,GeV. 

We also extend the fermion sector of the GG model with one $15$-dimensional vectorlike set of fields, where these additional fermions are denoted as $15_F+\overline{15}_F \equiv \Sigma+\overline{\Sigma}$ with $15_F\equiv \Sigma=\Sigma_1(1,3,1)+  \Sigma_3(3,2,1/6)+\Sigma_6(6,1,-2/3)$. The SM fermions reside in ${\overline{5}_F}_i$ and ${10_F}_i$, where $i(=1,2,3)$ is the generation index. The scalar and fermion representations of our proposal thus comprise $5_H$, $24_H$, $35_H$, ${\overline{5}_F}_i$, ${10_F}_i$, and $15_F+\overline{15}_F$, where $i=1,2,3$. 

\subsection{Gauge coupling unification}

We want to demonstrate that the model generates gauge coupling unification at the one-loop level. The introduction of $35_H$ and $15_F+\overline{15}_F$ is crucial in that regard since the SM multiplets in $24_H$ and $5_H$ cannot produce viable unification on their own. There are, however, two mass relations between the multiplets in $35_H$ and in $15_F+\overline{15}_F$ that one needs to take into account in order to perform a proper unification analysis as we discuss next.

It is well known that the masses of $\phi_1 \in 24_H$ and $\phi_8 \in 24_H$ can be treated as arbitrary parameters, as far as the gauge coupling unification is concerned, since the potential contains enough nontrivial contractions of $24_H$ with itself to allow for that to happen. The model can also accommodate the splitting between the masses of $\Lambda_1 \in 5_H$ and $\Lambda_3 \in 5_H$, where we take that $M_{\Lambda_3} \geq 3 \times 10^{11}$\,GeV in order to satisfy experimental bounds on the partial proton decay lifetimes~\cite{Dorsner:2012uz} while the mass of $\Lambda_1$ is the mass of the SM Higgs field, i.e., $M_{\Lambda_1} \equiv M_H$. 

The interaction of $15_F+\overline{15}_F$ with $24_H$ can induce mass splitting between fermions in $15$-dimensional representation, which is a rank-2 symmetric tensor. Namely, the two contractions    
\begin{equation}
\label{eq:lagrangian_sigma}
\mathcal{L} \supset M_{\Sigma} \Sigma^{\alpha \beta} \overline{\Sigma}_{\alpha \beta}  +y \Sigma^{\alpha \beta} \phi_\beta^\gamma \overline{\Sigma}_{\alpha \gamma}\,,
\end{equation}
where $\alpha,\beta,\gamma(=1,2,3,4,5)$ are the $SU(5)$ indices, yield
\begin{align}
&M_{\Sigma_1}=M_{\Sigma}+\frac{y}{2} \sqrt{\frac{3}{5}} v_{24}\,,\\
&M_{\Sigma_3}=M_{\Sigma}+\frac {y} {4\sqrt {15}} v_{24}\,,\\
\label{eq:mass_relation_a}
&M_{\Sigma_6}=2 M_{\Sigma_3}-M_{\Sigma_1}.
\end{align}
Even though $\Sigma_3$ and $\Sigma_1$ mix with the SM fermions, as we show later, Eq.~\eqref{eq:mass_relation_a} holds in the particular part of parameter space we are interested in and we accordingly use it in the one-loop gauge coupling unification analysis. In fact, the preferred unification scenario corresponds to the case when the masses of $\Sigma_1$, $\Sigma_3$, and $\Sigma_6$ are degenerate for all practical purposes. We note that $y$ of Eq.~\eqref{eq:lagrangian_sigma} is, strictly speaking, Yukawa coupling that, in the case of degenerate masses of $\Sigma_1$, $\Sigma_3$, and $\Sigma_6$, does not affect mechanisms that yield correct values of the SM fermion masses and mixing parameters.

The masses of multiplets in $35_H$, which is a rank-3 completely symmetric tensor, are determined by the following interactions 
\begin{equation}
\mathcal{L} \supset \mu^2_{35}\Phi\Phi^{\ast}+\lambda_0 \left(\Phi\Phi^{\ast}\right)\phi^2
+\lambda_1 \Phi^{\alpha\beta\gamma}{\Phi^{\ast}}_{\alpha \delta \epsilon}\phi^\delta_{\beta}\phi^\epsilon_{\gamma} 
+\lambda_2 \Phi^{\alpha\beta \epsilon}{\Phi^{\ast}}_{\alpha\beta \delta}\phi^\gamma_\epsilon\phi^\delta_\gamma+\mu^{\prime}_{35}\Phi^{\alpha\beta\gamma}{\Phi^{\ast}}_{\delta\alpha\beta}\phi^{\delta}_{\gamma}\,,
\end{equation}
and the squares of the masses read
\begin{align}
&M^2_{\Phi_1}=\mu ^2_{35}+\frac{\lambda_0}{2}v_{24}^2+\frac{3\lambda _1}{20}
   v_{24}^2+\frac{3\lambda _2}{20} v_{24}^2+\frac{1}{2} \sqrt{\frac{3}{5}}
    \mu '_{35} v_{24}\,,\\
&M^2_{\Phi_{3}}=\mu ^2_{35}+\frac{\lambda_0}{2}v_{24}^2-\frac{\lambda _1}{60} 
   v_{24}^2+\frac{11 \lambda _2}{90} v_{24}^2+\frac{2 }{3
   \sqrt{15}} \mu '_{35}v_{24}\,,\\
&M^2_{\Phi_{6}}=\mu ^2_{35}+\frac{\lambda_0}{2}v_{24}^2-\frac{2 \lambda _1}{45}
   v_{24}^2+\frac{17 \lambda _2}{180}  v_{24}^2-\frac{1 }{6
   \sqrt{15}}\mu '_{35}v_{24}\,,\\
\label{eq:mass_relation_b}
&M^2_{\Phi_{10}}=M^2_{\Phi_1}-3M^2_{\Phi_3}+3 M^2_{\Phi_6}\,.
\end{align}
The last equality is the other mass relation that we need to use in our unification analysis.

To summarize, relevant degrees of freedom for the gauge coupling unification considerations are masses of $\Phi_1, \Phi_3,\Phi_6,\Phi_{10} \in 35_H$, $\Sigma_1,\Sigma_3,\Sigma_6 \in 15_F+\overline{15}_F$, $\phi_1,\phi_8 \in 24_H$, and $\Lambda_3 \in 5_H$ and the constraints that need to be satisfied are given by Eqs.~\eqref{eq:mass_relation_a} and~\eqref{eq:mass_relation_b} with $M_{\Lambda_3} \geq 3 \times 10^{11}$\,GeV. The gauge coupling unification is successful if and when $M_\mathrm{GUT}\geq 5\times 10^{15}$\,GeV, where $M_\mathrm{GUT}$ is the scale of unification, i.e., the GUT scale, and, at the same time, the mass of the proton decay mediating $SU(5)$ gauge bosons. The aforementioned limit on $M_\mathrm{GUT}$ is due to experimental constraint on the gauge-mediated $d=6$ proton decay.

\begin{figure}[th!]
\centering
\includegraphics[scale=1.4]{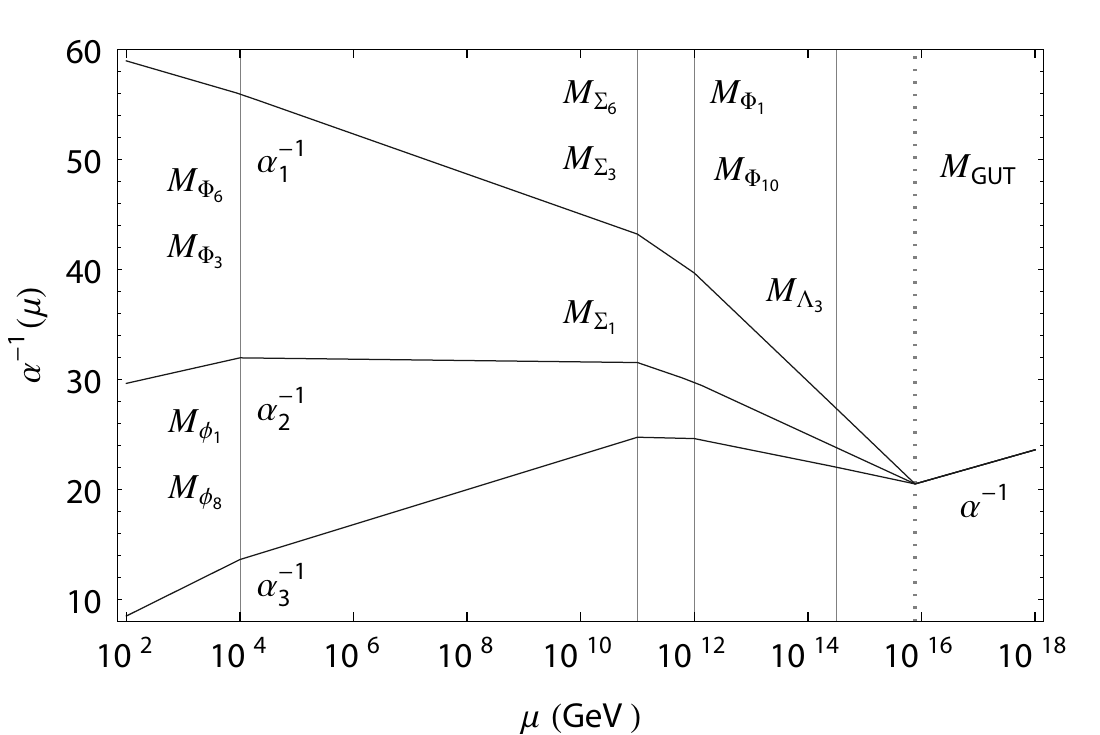}
\caption{The gauge coupling unification, at the one-loop level, that corresponds to the case when $M_{\Sigma_1}=10^{11}$\,GeV and $M_{\Phi_1}=10^{12}$\,GeV that yields correct neutrino mass scale.}
\label{fig:unification_a}
\end{figure}

The gauge coupling unification analysis reveals that the field $\Phi_1(1,4,-3/2)$ ($\Sigma_1(1,3,1)$) prefers to be very heavy (light) if $M_\mathrm{GUT}$ is to be sufficiently large. But, for the reasons that will become clear when we discuss the neutrino mass generation mechanisms, the masses of $\Phi_1$ and $\Sigma_1$ should be of the same order of magnitude if one is to have potentially viable scenario. Also, $\Phi_1$ cannot be too heavy since the neutrino mass scale would not come out right. Due to these conflicting needs we are left with rather limited parameter space where one can simultaneously address experimental results on proton decay lifetimes and neutrino masses in a proper manner. We accordingly present, in Fig.~\ref{fig:unification_a}, the gauge coupling unification scenario that corresponds to the case when $M_{\Sigma_1}=10^{11}$\,GeV and $M_{\Phi_1}=10^{12}$\,GeV that yields $M_\mathrm{GUT}=7.7 \times 10^{15}$\,GeV under the assumption that the lightest new physics states cannot reside below $10$\,TeV. We use that particular unification scenario in our numerical study of the fermion masses and mixing parameters to showcase viability of our proposal. The vertical lines in Fig.~\ref{fig:unification_a} correspond to the mass scales of the relevant multiplets that we explicitly specify for clarity of exposition. 

To generate Fig.~\ref{fig:unification_a} we use $\alpha_S(M_Z)=0.1193$, $\alpha^{-1}(M_Z)=127.906$, and
$\sin^2 \theta_W=0.23126$~\cite{Agashe:2014kda} to maximise the GUT scale for the case when we set $M_{\Sigma_1}=10^{11}$\,GeV and $M_{\Phi_1}=10^{12}$\,GeV while the masses of all other fields are allowed to freely move between $10$\,TeV and $M_\mathrm{GUT}$ except for $\Lambda_3$ multiplet that needs to have $M_{\Lambda_3} \geq 3 \times 10^{11}$\,GeV. Again, all this is done after we implement relations given in Eqs.~\eqref{eq:mass_relation_a} and~\eqref{eq:mass_relation_b}. 

\begin{figure}[t!]
\centering
\includegraphics[scale=1.4]{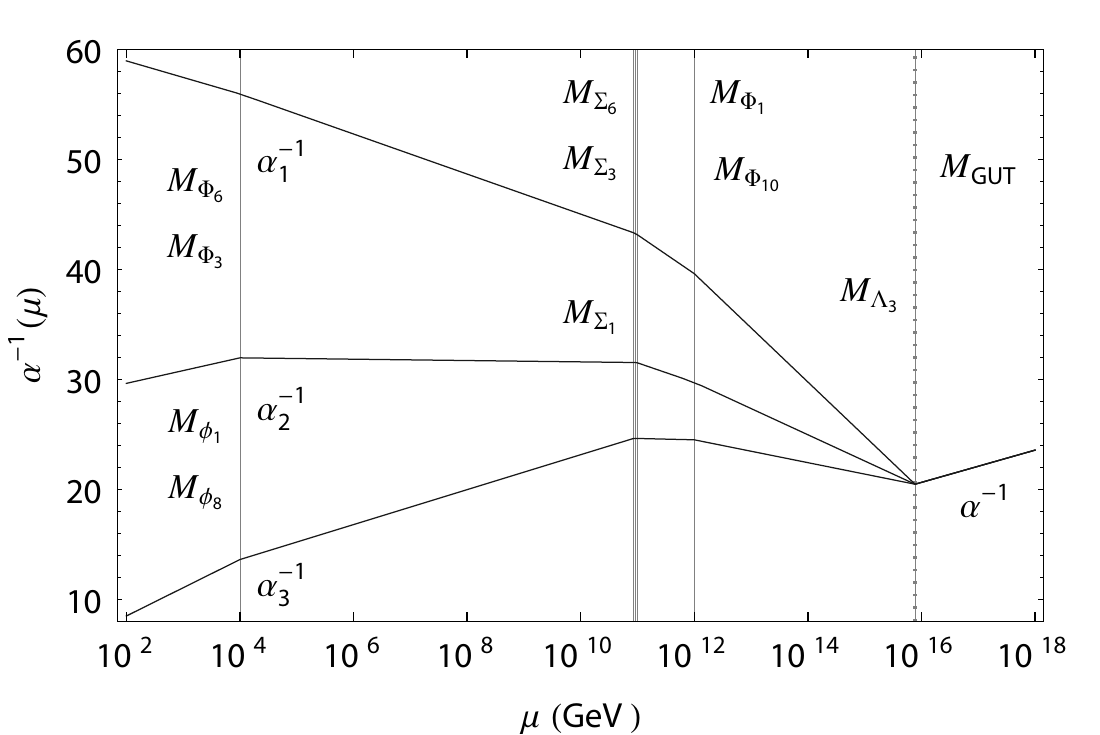}
\caption{The gauge coupling unification, at the one-loop level, that corresponds to the case when $M_{\Sigma_1}=10^{11}$\,GeV, $M_{\Sigma_3}=9.2 \times 10^{10}$\,GeV, $M_{\Sigma_6}=8.4 \times 10^{10}$\,GeV, $M_{\Phi_1}=10^{12}$\,GeV, and $M_{\Lambda_3}=M_\mathrm{GUT} = 7.7 \times 10^{15}$\,GeV.}
\label{fig:unification_b}
\end{figure}

It is clear from Fig.~\ref{fig:unification_a} that the proposal yields several multiplets with nontrivial assignments under $SU(3)$ and/or $SU(2)$ that could potentially reside at the TeV scale. These multiplets are $\Phi_6$, $\Phi_3$, $\phi_8$, and $\phi_1$. Note that in the regime of interest, i.e., when $M_{\Sigma_1}$ is comparable to $M_{\Phi_1}$, the masses of fermions in $15_F+\overline{15}_F$ are degenerate for all practical purposes. The degeneracy of states in $15_F+\overline{15}_F$ that demands $y \approx 0$ is not guaranteed by any symmetry. It is  simply an outcome of our proposal when the gauge coupling unification requirement and proton decay constraints are combined with neutrino oscillation data. Also, this unification scenario implies that the $d=6$ proton decay contributions due to scalar mediation are completely negligible since $\Lambda_3$ is the only scalar leptoquark in the proposed model, where the maximisation of the GUT scale yields $M_{\Lambda_3} \geq 3.1 \times 10^{14}$\,GeV. If we set $M_{\Sigma_1}=10^{11}$\,GeV and $M_{\Phi_1}=10^{12}$\,GeV as in the unification scenario presented in Fig.~\ref{fig:unification_a} but, this time around, insist that $\Lambda_3$ is at the GUT scale we obtain the unification of gauge couplings shown in Fig.~\ref{fig:unification_b}. The impact of $\Lambda_3$ on the gauge coupling running is completely compensated by a tiny split between the masses of $\Sigma_1$, $\Sigma_3$, and $\Sigma_6$ with $y=1.2 \times 10^{-6}$, whereas the GUT scale is marginally larger than in the scenario shown in Fig.~\ref{fig:unification_a} but rounds to the same numerical value.  

Finally, we note that the multiplet $\Phi_1(1,4,-3/2)$ cannot be below $10^{11}$\,GeV ($10^{10}$\,GeV) under the assumption that the lightest new physics states introduced in this proposal do not reside below $10$\,TeV ($1$\,TeV) if the GUT scale is to exceed $5 \times 10^{15}$\,GeV.

\subsection{Neutrino mass generation}
\begin{figure}[t!]
\centering
\includegraphics[width=1\textwidth]{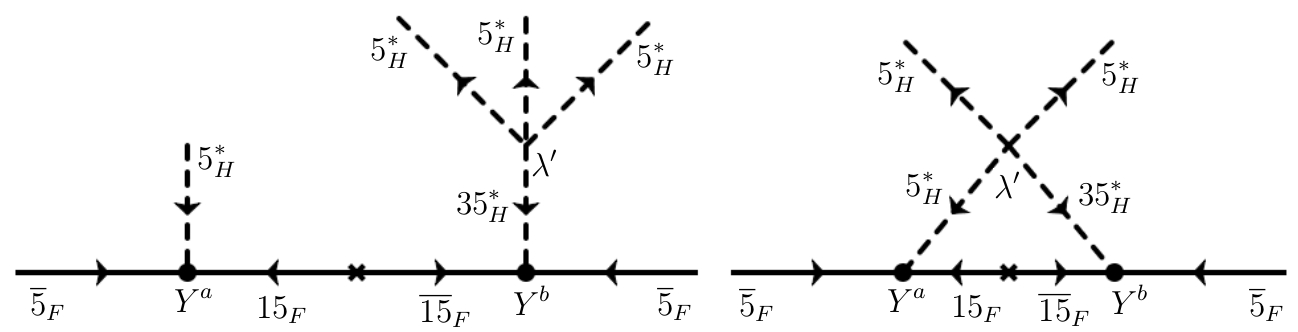}
\caption{The Feynman diagrams of the tree level (left panel) and the one-loop level (right panel) contributions towards Majorana neutrino masses.}
\label{fig:diagram}
\end{figure}

The $SU(5)$ contractions that generate contributions towards Majorana neutrino masses read
\begin{equation}
\label{eq:lagrangian_neutrino}
\mathcal{L}\supset \lambda^{\prime}\; 5_H5_H5_H35_H+Y^a_i\;15_F{\overline{5}_F}_i5_H^{\ast}+Y^{b}_i\;\overline{15}_F{\overline{5}_F}_i35_H^{\ast}\,,    
\end{equation}
where $Y^a$ and $Y^b$ are arbitrary $1 \times 3$ Yukawa coupling matrices and $\lambda^{\prime}$ is a dimensionless parameter. The neutrino mass contributions are generated both at the tree level via $d=7$ operator and at the one-loop level via $d=5$ operator~\cite{Babu:2009aq}. The Feynman diagram for the former (latter) contribution is shown in the left (right) panel of Fig.~\ref{fig:diagram}.

We obtain the following contributions towards neutrino mass matrix elements: 
\begin{align}
&(\mathcal{M}_{\nu}^{d=7})_{ij}=-\lambda^{\prime}\frac{v^4_H}{M_{\Sigma_1}M^2_{\Phi_1}} (Y^a_iY^b_j+Y^b_iY^a_j)\,,
\label{eq:d7}
\\
&(\mathcal{M}^{d=5}_{\nu})_{ij}= \frac{\lambda^{\prime}v_H^2}{16 \pi^2} 
\frac{(Y^a_iY^b_j+Y^b_iY^a_j) M_{\Sigma_1}}{M^2_{\Phi_1}-M^2_H}
\left( 
\frac{M^2_{\Phi_1}}{M^2_{\Sigma_1}-M^2_{\Phi_1}}
\log \left( \frac{M^2_{\Sigma_1}}{M^2_{\Phi_1}} \right)
-
\frac{M^2_{H}}{M^2_{\Sigma_1}-M^2_{H}}
\log \left( \frac{M^2_{\Sigma_1}}{M^2_{H}} \right) \right)\,,
\label{eq:d5}
\end{align}
where both $\mathcal{M}_{\nu}^{d=7}$ and $\mathcal{M}_{\nu}^{d=5}$ exhibit exactly the same flavor dependence.

In the regime of interest, i.e., when $M_{\Sigma_1},  M_{\Phi_1} \gg M_H, v_H$, it is the $d=5$ contribution that dominates and we find that the entries of the neutrino mass matrix ($\mathcal{M}_{\nu} \equiv \mathcal{M}^{d=5}_{\nu}$) are 
\begin{align}
(\mathcal{M}_{\nu})_{ij}&\sim \frac{\lambda^{\prime}v_H^2}{16 \pi^2} (Y^a_iY^b_j+Y^b_iY^a_j) \frac{M_{\Sigma_1}}{M^2_{\Sigma_1}-M^2_{\Phi_1}}
\log \left( \frac{M^2_{\Sigma_1}}{M^2_{\Phi_1}} \right)  = m_0 (Y^a_iY^b_j+Y^b_iY^a_j)\,,
\end{align}
where, for $\lambda^{\prime}=1$, $M_{\Sigma_1} = 10^{11}$\,GeV, and $M_{\Phi_1} = 10^{12}$\,GeV, one finds that $m_0 \approx 4 \times 10^{-11}$\,GeV. This, in turn, is the right neutrino mass scale if the entries in $Y^a$ and $Y^b$ are $\mathcal{O}(1)$ parameters. The tree level contribution for the unification scenario under consideration is very suppressed compared to the one-loop level one. Namely, we find that the prefactor of the term $(Y^a_iY^b_j+Y^b_iY^a_j)$ in Eq.~\eqref{eq:d7} is  
$\lambda^{\prime}(v^4_H)/(M_{\Sigma_1}M^2_{\Phi_1}) \approx 10^{-26}$\,GeV. Note that one neutrino is always massless in our proposal since $\det(\mathcal{M}_{\nu})=0$. 

\subsection{Charged fermion masses}
The contractions that are relevant for the charged fermion mass generation read
\begin{align}
\mathcal{L} \supset &Y^d_{ij}{10_F}_i{\overline{5}_F}_j5^{\ast}_H+ Y^u_{ij}   {10_F}_i{10_F}_j5_H +Y^{c}_{i} {10_F}_i\overline{15}_F24_H
\nonumber \\ &
+Y^{a}_{i} 15_F{\overline{5}_F}_i5^{\ast}_H+M_{\Sigma}\overline{15}_F15_F+y\; \overline{15}_F15_F 24_H\,.
\label{lag}
\end{align}
Here, we use the freedom to freely rotate ${10_F}_i$ and ${\overline{5}_F}_j$ in an  independent manner to go into the basis where $Y^d$ is a real and diagonal matrix. $Y^u$ is a symmetric $3 \times 3$ complex matrix, whereas $Y^{c}$ is a $3 \times 1$ complex matrix. The contraction proportional to the matrix $Y^{a}$ also appears in Eq.~\eqref{eq:lagrangian_neutrino} in regard to the discussion of the neutrino mass generation while the last two terms in Eq.~\eqref{lag} appear in Eq.~\eqref{eq:lagrangian_sigma} and are relevant for the mass splitting of the multiplets within $15_F+\overline{15}_F$. These facts simply mean that there is a direct connection between the mass generation mechanisms for the charged and neutral fermions in our proposal. In fact, the role of $15_F+\overline{15}_F$ is to simultaneously generate neutrino masses and induce an experimentally observed mismatch between the masses of the charged leptons and the down-type quarks as we discuss next. Note that the possibility to address the aforementioned mismatch with a vectorlike $15$-dimensional representation within the context of supersymmetric $SU(5)$ has been studied in Ref.~\cite{Oshimo:2009ia}. 

The decomposition of ${10_F}_i$ and ${\overline{5}_F}_i$ under the SM gauge group is ${10_F}_i={Q_L}_i(3,2,1/6)+u^c_i(\overline{3},1,-2/3)+e^c_i(1,1,1)$ and ${\overline{5}_F}_i={\ell_L}_i(1,2,-1/2) +d^c_i(\overline{3},1,-1/3)$, respectively. After the electroweak symmetry breaking the isodoublet (isotriplet) fermions decompose under the $SU(3) \times U(1)_\mathrm{em}$ group as ${Q_L}_i=u_i(3,2/3)+d_i(3,-1/3)$ and $\Sigma_3=\Sigma^u(3,2/3)+\Sigma^d(3,-1/3)$ ($\Sigma_1=\Sigma^{\nu}(1,0)+\Sigma^{e^c}(1,1)+ \Sigma^{e^ce^c}(1,2)$), where the second number in the parentheses represents electric charge in units of absolute value of the electron charge. The mixing between the fermions in ${10_F}_i$ and $15_F+\overline{15}_F$ that appears as a result of the $SU(5)$ symmetry breaking reads 
\begin{align}
\mathcal{L} \supset -\frac{1}{4}\sqrt{\frac{5}{3}}v_{24} Y^c_i\; {Q_L}_i\overline{\Sigma}_3\,,
\end{align}
whereas the electroweak symmetry breaking induces the following mixings among the fermions: 
\begin{align}
\mathcal{L} \supset &
\begin{pmatrix}u_i&\Sigma^u\end{pmatrix}
\begin{pmatrix}
v_H\left(Y^u_{ij}\right)_{3\times 3}&-\frac{1}{4}\sqrt{\frac{5}{3}}v_{24}\left(Y^c_i\right)_{3\times 1}\\
\left(0\right)_{1\times 3}&\left(M_{\Sigma_3}\right)_{1\times 1}
\end{pmatrix}
\begin{pmatrix}u^c_j\\\overline{\Sigma}^u\end{pmatrix}
\nonumber \\&
+\begin{pmatrix}d_i&\Sigma^d\end{pmatrix}
\begin{pmatrix}
v_H^{\ast}\left(Y^d_{ij}\right)_{3\times 3}&-\frac{1}{4}\sqrt{\frac{5}{3}}v_{24}\left(Y^c_i\right)_{3\times 1}\\
v_H^{\ast} \left(Y^a_j\right)_{1\times 3}&\left(M_{\Sigma_3}\right)_{1\times 1}
\end{pmatrix}
\begin{pmatrix}d^c_j\\\overline{\Sigma}^d\end{pmatrix}
\nonumber \\&
+\begin{pmatrix}e_i&\overline{\Sigma}^{e^c}\end{pmatrix}
\begin{pmatrix}
v_H^{\ast}\left({Y^d}^T_{ij}\right)_{3\times 3}&v_H^{\ast}\left(Y^a_i\right)_{3\times 1}\\ \left(0\right)_{1\times 3}&\left(M_{\Sigma_1}\right)_{1\times 1}
\end{pmatrix}
\begin{pmatrix}e^c_j\\\Sigma^{e^c}\end{pmatrix}.
\end{align}
The states $\Sigma^{u,d,e^c}+\overline{\Sigma}^{u,d,e^c}$ need to be very heavy if one is to generate the correct neutrino mass scale, and can be safely integrated out. We accordingly find, in the limit when $v_{24} Y^c, M_{\Sigma_{1,3}} \gg v_H$, that the mass matrices for the up-type quarks ($M_u$), down-type quarks ($M_d$), and charged leptons ($M_e$) are 
\begin{align}
&M_u=\left( \mathbb{I}_{3\times 3}+\delta^2\;Y^c{Y^c}^{\dagger} \right)^{-\frac{1}{2}} v_H Y^u,  \label{massu}
\\
&M_d=\left( \mathbb{I}_{3\times 3}+\delta^2\;Y^c{Y^c}^{\dagger} \right)^{-\frac{1}{2}} v_H \left( Y^d + \delta\; Y^cY^a  \right),  \label{massd}
\\
&M_e=v_H{Y^d}^T \label{masse},
\end{align}
where we define the dimensionless parameter $\delta \equiv \sqrt{5/3} v_{24}/(4 M_{\Sigma_3})$ and, again, take $v_H$ to be real. The gauge coupling unification scenario we present in Fig.~\ref{fig:unification_a} yields $\delta=2.49\times 10^4$. Note that without the mixing one retrieves the GG prediction $M_e=M_d^T$, at the unification scale, which is in conflict with experimental observations after the measured values of the down-type quarks and charged leptons are evolved to the GUT scale.

Again, the neutrino mass matrix elements are
\begin{align}
(\mathcal{M}_{\nu})_{ij}
&= m_0 \left(
Y^a_i Y^b_j+Y^b_i Y^a_j
\right)= (U_\mathrm{PMNS}\;
 \mathrm{diag}(m_1,m_2,m_3)\;
U_\mathrm{PMNS}^T)_{ij}\,, 
\label{nu}
\end{align}
where $m_i$, $i=1,2,3$, are neutrino mass eigenstates and $U_\mathrm{PMNS}$ represents the Pontecorvo-Maki-Nakagawa-Sakata (PMNS) unitary mixing matrix. Since we work in the basis where $Y^d$ is diagonal and real, the unitary matrix that diagonalizes the neutrino matrix must be identified with the PMNS matrix. Finally, the masses of the heavy states are
\begin{align}
\label{eq:heavy}
&M_{\Sigma^u}=M_{\Sigma_3}\left( 1+\delta^2\; {Y^c}^{\dagger}Y^c \right)^{\frac{1}{2}},\;\; 
M_{\Sigma^d}=M_{\Sigma^u},\;\; 
M_{\Sigma^{e^c}}=M_{\Sigma_1}.
\end{align}

To summarize, our proposal uses one vectorlike $15$-dimensional representation that simultaneously generates neutrino masses with the aid of one $35$-dimensional scalar representation and creates a viable mismatch between the masses of the down-type quarks and charged leptons. There are only two (three) $3 \times 3$ ($3 \times 1$) Yukawa coupling matrices $Y^u$ and $Y^d$ ($Y^a$, $Y^b$, and $Y^c$) to accommodate the experimentally measured fermion masses and mixing parameters. Moreover, without the loss of generality, one can go into a basis where $Y^d$ is a diagonal and real matrix while $Y^u$ is a symmetric matrix. Note that the dimensionless parameter $y$ is, strictly speaking, Yukawa coupling, but the unification considerations imply that it can be neglected for all practical purposes. 

\subsection{Numerical analysis}

In this section, we perform a numerical fit of the SM fermion masses and mixing parameters that corresponds to the unification scenario of Fig.~\ref{fig:unification_a} to demonstrate viability of our proposal. The fermion mass matrices are given in Eqs.~\eqref{massu},~\eqref{massd},~\eqref{masse}, and~\eqref{nu}. Since the charged lepton mass matrix is already diagonal in our basis, i.e., $M_e=v_H\;\mathrm{diag}(y_e,y_{\mu},y_{\tau})$, we can trivially determine the entries of the $Y^d(=(y_e,y_{\mu},y_{\tau})=(2.703\times 10^{-6}, 5.707\times 10^{-4}, 9.70\times 10^{-3}))$ matrix using the GUT-scale values of the observables that are listed in Table~\ref{CHARGE}. 

Since the down-type quark and neutrino mass matrices share the same Yukawa couplings $Y^a_i$, $i=1,2,3$, we perform a combined fit of these two sectors to reproduce the correct down-type quark masses and neutrino observables. We obtain the following fit parameters: 
\begin{align}
Y^a&=(-0.0899, 0.551, 1),\label{fit1}\\
Y^b&=(0.975, 2.381, 1),\label{fit2}\\
Y^c&=-1.865 \times 10^{-7} (0.00137,  0.0942, 1),\label{fit3}
\end{align}
where we normalize $Y^a_3$ and $Y^b_3$ to 1. To be consistent with our unification scenario and this normalization, the overall scale for the neutrino mass matrix is fixed to be $m_0=9.28 \times 10^{-12}$\,GeV which, in turn, requires $\lambda^{\prime}=0.239$.
We summarize the best-fit values in Tables \ref{CHARGE} and \ref{NU}. Clearly, all the observables can be fitted to their experimentally measured central values given in Tables \ref{CHARGE} and \ref{NU}, while one neutrino is predicted to be massless. 

In general the Yukawa couplings in $Y^{a,b,c}$ are complex numbers, but for simplicity we have taken them to be real.  Note that the up-type quark mass matrix is proportional to the complex symmetric matrix $Y^u$. This provides enough freedom for one to simultaneously reproduce the up-type quark masses and the Cabibbo-Kobayashi-Maskawa (CKM) parameters. It should be pointed out that to fit the CKM parameters one needs to keep track of the unitary transformations that take the down-type quark mass matrix given in Eq.~\eqref{massd} into a diagonal form. Note that the fit implies that $(\delta\; Y^c_i) \ll 1$, where, again, $\delta \equiv \sqrt{5/3} v_{24}/(4 M_{\Sigma_3})$. This means that the mass relation of Eq.~\eqref{eq:mass_relation_a} is not affected by the fermion mixing mechanism, as can be seen from Eq.~\eqref{eq:heavy}. More importantly, it means that the up-type quark mass matrix, given in Eq.~\eqref{massu}, is a symmetric one to great accuracy. Our proposal thus predicts that there are two proton decay channels that depend only on the CKM parameters and the unification input~\cite{FileviezPerez:2004hn} of Fig.~\ref{fig:unification_a}. These channels are $p \rightarrow K^+ \overline{\nu}$ and $p \rightarrow \pi^+ \overline{\nu}$.

\begin{table}[th!]
\centering
\footnotesize
\resizebox{0.5\textwidth}{!}{
\begin{tabular}{|c|c|c|}\hline
\multicolumn{3}{ |c| }{Down-type quark \;\&\; charged lepton masses}
\\ 
\cline{1-3}
Observable &  Input (GeV) & Fit (GeV)  \\ [1ex] \hline
$m_{d}/10^{-3}$  &1.14$\pm$0.11&1.14  \\ \hline
$m_{s}/10^{-2}$  &2.15$\pm$0.11&2.15 \\ \hline
$m_{b}$  &0.99$\pm$0.009 &0.99 \\ \hline
$m_{e}/10^{-4}$  &4.707$\pm$0.0047&4.707 \\ \hline
$m_{\mu}/10^{-2}$  &9.936$\pm$0.0099&9.936 \\ \hline
$m_{\tau}$  &1.689$\pm$0.0016&1.689\\ \hline
\end{tabular}}
\caption{The fit input for the GUT scale values of the down-type quark and charged lepton masses~\cite{Babu:2016bmy} and the corresponding fit output.}\label{CHARGE}
\end{table}

\FloatBarrier
\begin{table}[th!]
\begin{minipage}{.5\linewidth}
\centering
\footnotesize
\resizebox{0.85\textwidth}{!}{
\begin{tabular}{|c|c|}\hline
\pbox{10cm}{$\nu$ parameters} &Input  \\ [1ex] \hline
$\Delta m^{2}_{21}/10^{-5}$ &7.57$\pm$0.18 \;(eV$^{2}$)  \\ \hline
$\Delta m^{2}_{31}/10^{-3}$ &2.50$\pm$0.03 \;(eV$^{2}$)  \\ \hline
$\sin^{2}\theta^{\rm{PMNS}}_{12}$  &0.322$\pm$0.018  \\ \hline
$\sin^{2}\theta^{\rm{PMNS}}_{23}$ &0.542$\pm$0.025  \\ \hline
$\sin^{2}\theta^{\rm{PMNS}}_{13}/10^{-2}$ &2.219$\pm$0.075\\ \hline
\end{tabular}}
\end{minipage}
\begin{minipage}{.5\linewidth}
\centering
\footnotesize
\resizebox{0.92\textwidth}{!}{
\begin{tabular}{|c|c|c|c|}
\hline
\pbox{10cm}{$\nu$ masses} &\pbox{10cm}{\vspace{3pt} Fit (eV)}  & \pbox{10cm}{$\nu$ angles} &\pbox{10cm}{\vspace{3pt} Fit ($^{\circ}$)\vspace{3pt}}  \\ [1ex] \hline
$m_1$&0&$\theta^{\rm{PMNS}}_{12}$& 34.57   \\ \hline
$m_2/10^{-3}$&8.70&$\theta^{\rm{PMNS}}_{23}$& 47.41    \\ \hline
$m_3/10^{-2}$&4.99&$\theta^{\rm{PMNS}}_{13}$& 8.56   \\ \hline
\end{tabular}}
\end{minipage}%
\caption{The fit input for the neutrino observables~\cite{deSalas:2017kay} and the corresponding fit output.}\label{NU}
\end{table}

\section{Conclusion}
\label{SEC-03}

We proposed a simple $SU(5)$ model that relates the neutrino mass generation mechanism to the observed disparity between the masses of charged leptons and down-type quarks. The entire structure of the model is based on the first five nontrivial $SU(5)$ representations of the lowest possible dimension. The scalar sector is made up of three representations of dimensions $5$, $24$, and $35$, while the fermion sector---besides the usual matter content of the minimal $SU(5)$---contains only one set of $15$-dimensional vectorlike fields. The role of the vectorlike fields is twofold. They generate neutrino masses and, at the same time, create required mismatch between the down-type quarks and charged leptons. The proposal successfully accommodates all experimentally measured fermion masses and mixing parameters with two (three) $3 \times 3$ ($3 \times 1$) Yukawa coupling matrices, where without loss of generality one of the $3 \times 3$ matrices can be taken to be diagonal and real, whereas the other $3 \times 3$ matrix is symmetric.

The gauge coupling unification considerations, coupled with phenomenological constraints inferred from experiments that probe neutrino masses and mixing parameters and/or look for proton decay, fix all relevant mass scales of the model. The proposed scenario places several multiplets with nontrivial assignments under $SU(3)$ and/or $SU(2)$ at the scales potentially accessible at the LHC and future colliders and correlates this feature with the gauge-boson mediated proton decay signatures. Two particular decay channels, i.e., $p \rightarrow K^+ \overline{\nu}$ and $p \rightarrow \pi^+ \overline{\nu}$, depend only on the scale of unification and the SM parameters. The model contains only one scalar leptoquark whose contribution towards proton decay is negligible. It also predicts the existence of one massless neutrino.

\section*{Acknowledgments}

I.D.\ acknowledges support of COST Action CA15108.

\FloatBarrier

\end{document}